\documentclass[12pt]{article}
\usepackage{amsmath}
\usepackage{amssymb}
\usepackage{epsfig}
\usepackage[latin5]{inputenc}

\numberwithin{equation}{section}

\newcommand{\gen}[1]{\partial_{#1}}

\newcommand{\curl}[1]{ \{#1\} }

\newcommand{\nil}{\mathfrak {h}}

\newcommand{\sch}{\mathfrak {sch}}

\newcommand{\dx}{\partial_x}

\newcommand{\dt}{\partial_t}

\newcommand{\dw}{\partial_{\omega}}
\newcommand{\dr}{\partial_{\rho}}

\DeclareMathOperator{\Sl}{sl}

\DeclareMathOperator{\So}{so}

\DeclareMathOperator{\simi}{\mathfrak{gs}}
\DeclareMathOperator{\gal}{\mathfrak{gal}}
\DeclareMathOperator{\gl}{gl}

\newtheorem{prop}{Proposition}
\newtheorem{thm}{Theorem}
\newtheorem{cor}{Corollary}
\newtheorem{rmk}{Remark}
\newtheorem{defn}{Definition}

\begin{document}
\title{\bf
\large Symmetry classification of variable coefficient cubic-quintic nonlinear Schr\"{o}dinger equations}

\author{C. \"{O}zemir\footnotemark[1]\thanks{Department of Mathematics, Faculty of Science and Letters,
Istanbul Technical University, 34469 Istanbul,
Turkey, e-mail: ozemir@itu.edu.tr} \and F. G\"{u}ng\"{o}r\thanks{Department of Mathematics, Faculty of Arts and Sciences, Do\u{g}u\c{s} University, 34722 Istanbul, Turkey, e-mail: fgungor@dogus.edu.tr}
}

\date{\today}

\maketitle

\begin{abstract}
A Lie-algebraic classification of the variable coefficient cubic-quintic nonlinear Schr\"{o}dinger  equations involving 5 arbitrary functions of space and time is performed under the action of equivalence transformations. It is shown that their symmetry group can be at most four-dimensional in the genuine cubic-quintic nonlinearity. It is only five-dimensional (isomorphic to the Galilei similitude algebra $\simi(1)$) when the equations are of cubic type, and six-dimensional (isomorphic to the Schr\"{o}dinger algebra $\sch(1)$) when they are of quintic type.
\end{abstract}



\section{Introduction}

In this paper we are interested in giving a classification of variable coefficient cubic-quintic nonlinear Schr\"{o}dinger (CQNLS) equations
\begin{equation}\label{cqsch}
iu_t+f(x,t)u_{xx}+k(x,t)\,u_x+g(x,t)|u|^2u+q(x,t)|u|^4u+h(x,t)u=0
\end{equation}
according to their Lie point symmetries up to equivalence.
 Here $u$ is a complex-valued function, $f$ is real-valued, $k, g, q, h$ are complex-valued functions of the form $k=k_1(x,t)+i\,k_2(x,t)$, $g(x,t)=g_1(x,t)+ig_2(x,t)$, $q(x,t)=q_1(x,t)+iq_2(x,t)$ and
$h(x,t)=h_1(x,t)+ih_2(x,t)$. We assume that  $g\not\equiv0$ or $q\not\equiv 0$, that is,
at least one of $g_1,g_2,q_1,q_2$ is different from zero. Eq. \eqref{cqsch} contains two physically important equations: cubic Schr\"{o}dinger equation for $k=q=0$ and quintic Schr\"{o}dinger equation for $k=g=0$ in one space dimension. Specifically, the case $k=(n-1)/x$ corresponds to the radial counterpart of the cubic and quintic equations in $n$ space dimensions with $x$ playing the role of the radial coordinate when $n\geq 2$ and real half-line when $n=1$.

Symmetry classification of \eqref{cqsch} in the special case $k=q=0$ was given in \cite{GagnonWinternitz1993}. Similar works based on a different approach appeared in \cite{PopovychIvanovaEshragi2004,R.Eshragi2010}. An in-depth analysis of the constant coefficient version of \eqref{cqsch} with $k=0$ in 3+1-dimensions was done in a series of papers \cite{GagnonWinternitz1988, GagnonWinternitz1989, GagnonGrammaticosRamaniWinternitz1989} where the authors studied Lie point symmetries and gave a complete subalgebra classification of symmetry algebras, reductions and  a comprehensive analysis of the explicit (group-invariant) solutions. Recently we looked at the solutions of the cubic version admitting only 4-dimensional Lie point symmetries \cite{Oezemir2011}. Other symmetry classification results relevant to several one and multidimensional versions of the nonlinear Schr\"{o}dinger equations involving arbitrary functions depending not only on space-time variables but also on dependent variables and its space derivatives can be found for example in \cite{Nikitin2001, NattermannDoebner1996, ZhdanovRoman2000}.

\section{Equivalence group and symmetries} By definition, the equivalence group of (\ref{cqsch}) is the group of transformations preserving the form of (\ref{cqsch}). This is given in the following definition.

\begin{defn} The equivalence group $\mathcal{E}$ of equation (\ref{cqsch}) is the group of smooth transformations $(t,x,u)\to (\tilde{t},\tilde{x},\tilde{u})$ preserving the differential structure. More precisely, $\mathcal{E}$  maps \eqref{cqsch} to
\begin{equation}\label{vcsd}
i\tilde{u}_{\tilde{t}}+\tilde{f}\tilde{u}_{\tilde{x}\tilde{x}}+\tilde{k}\tilde{u}_{\tilde{x}}+\tilde{g}|\tilde{u}|^2\tilde{u}
+\tilde{q}|\tilde{u}|^4\tilde{u}+\tilde{h}\tilde{u}=0.
\end{equation}
The equivalence group leaves the differential terms invariant but changes the coefficients, namely it leaves the equation form-invariant. A point symmetry group is a subgroup of $\mathcal{E}$ obtained when the coefficients remain unchanged under $\mathcal{E}$. These transformations are sometimes called allowed or admissible  transformations.
\end{defn}
Two approaches can be taken to find $\mathcal{E}$. One is the infinitesimal method and requires solving a large system of overdetermined linear partial differential equations just like determining the infinitesimal symmetries. The disadvantage of this approach is that discrete equivalence group does not come up as a subgroup. The other  is the direct approach and will be used below.

\begin{prop}
The equivalence group $\mathcal{E}$ of equation (\ref{cqsch}) is given by
\begin{equation}\label{equiv}
\mathcal{E}: \quad \tilde{t}=T(t),\quad \tilde{x}=X(x,t),\quad u=Q(x,t)\tilde{u},
\end{equation}
where the coefficients transform by
\begin{subequations}
\begin{eqnarray}\label{f}
&&\tilde{f}=\frac{f X_x^2}{\dot{T}},\\\label{g}
&&\tilde{g}=\frac{g|Q|^2}{\dot{T}},\\\label{q}
&&\tilde{q}=\frac{q|Q|^4}{\dot{T}},\\\label{h}
&&\tilde{h}=\frac{1}{\dot{T}}\left(h+i
\frac{Q_t}{Q}+f\frac{Q_{xx}}{Q}+k\frac{Q_x}{Q}\right),\\\label{k}
&&\tilde{k}=\frac{1}{\dot{T}}\left(iX_t+fX_{xx}+2 f
X_x\frac{Q_x}{Q}+kX_x\right)
\end{eqnarray}
\end{subequations}
under the conditions $X_x\ne 0$, $\dot{T}\ne 0$, $Q\ne 0$.
\end{prop}
For the sake of convenience we introduce the following moduli and phases for the complex functions $Q$, $u$ and $\tilde{u}$
$$Q(x,t)=R(x,t)e^{i\theta(x,t)},\quad u=\rho(x,t)\,
e^{i\omega(x,t)},\quad \tilde{u}=\sigma(\tilde{x},\tilde{t})\, e^{i\phi(\tilde{x},\tilde{t})}.$$
\subsection{Canonical cubic-quintic nonlinear Schr\"{o}dinger equation}
We can use the equivalence group to transform \eqref{cqsch} to some canonical form by choosing the free functions in the transformation suitably. Indeed, first of all, one can normalize $f\to 1$ by restricting $X$ to
\begin{equation}\label{x}
X(x,t)=\epsilon \sqrt{\dot{T}}x+\xi(t), \quad \epsilon=\mp1.
\end{equation}
With this choice of $X$, $\tilde{k}$ can be made zero by taking $R$ and $\theta$ as solutions of the following equations
\begin{equation}
2\frac{R_x}{R}+k_1=0,\quad
2X_x\theta_x+X_t+k_2X_x=0.
\end{equation}
So the canonical form of \eqref{cqsch} can be written in the form
\begin{equation}\label{canonic}
iu_t+u_{xx}+g(x,t)|u|^2u+q(x,t)|u|^4u+h(x,t)u=0.
\end{equation}
\begin{cor}
The equivalence group of the canonical equation \eqref{canonic} is
\begin{equation}\label{equivc}
\mathcal{E}:\quad \tilde{t}=T(t), \quad x=\epsilon \sqrt{\dot{T}}x+\xi(t), \quad \epsilon=\mp1,\quad
u=R_0(t)e^{i\theta(x,t)}\tilde{u},
\end{equation}
where
$$\dot{T}\ne 0, \quad R_0(t)\ne 0, \quad \theta(x,t)=-\frac{\ddot{T}}{8\dot{T}}\;x^2-\frac{\dot{\xi}}{2\epsilon
\sqrt{\dot{T}}}\,x+\eta(t),$$ and the transformed new coefficients are
\begin{align}\label{gqh}
\tilde{g}=\frac{gR_0^2}{\dot{T}}, \quad
\tilde{q}=\frac{qR_0^4}{\dot{T}},  \quad
\tilde{h}=\frac{1}{\dot{T}}\Big[(h_1-\theta_t-\theta_x^2+\Big(h_2+\frac{\dot{R_0}}{R_0}+\theta_{xx}\Big)\Big].
\end{align}
Notice that since $\theta$ is a second degree polynomial in $x$ and $R_0$ depends on only $t$, not every potential function $h$ can be killed through these transformations.
We recall the following relations
\begin{equation}
\sigma(\tilde{x},\tilde{t})=\frac{\rho(x,t)}{R_0(t)}, \quad
\phi(\tilde{x},\tilde{t})=\omega(x,t)-\theta(x,t).
\end{equation}
\end{cor}

\begin{rmk}
If $g$ and $q$ have no dependence on $x$ we can set $g_1\to 1$ and $q_1\to 1$ by a reparametrization of time.
\end{rmk}

\section{Symmetry group and determining equations}
The symmetry algebra is generated by vector fields of the form
$$Q=\xi(x,t,\rho,\omega)\gen x+\tau(x,t,\rho,\omega)\gen t+
\Phi(x,t,\rho,\omega)\gen \rho+\Psi(x,t,\rho,\omega)\gen \omega.$$
 The standard Lie infinitesimal algorithm requires calculating the second order prolongation of $Q$ to the jet space $J^2(\mathbb{R}^2,\mathbb{R}^2)$ with local coordinates $(t,x,\rho,\omega)$ and derivatives up to and including second order. The vector field $Q$ becomes an infinitesimal symmetry of the equation when its 2nd order prolongation annihilates  equation \eqref{canonic} written as a system in terms of $\rho, \omega$ on its solution manifold.
 The symmetry criterion provides  a set of over-determined linear partial differential equations. Among them solving those not involving coefficients we obtain the following assertion.
 \begin{prop}
 The symmetry
 algebra of the canonical CQNLS equation is generated by
the vector field
\begin{equation}\label{vf}
Q=\chi(x,t)\dx+\tau(t)\dt+A(t)\rho\dr+D(x,t)\dw,
\end{equation}
where the functions $\chi$ and $D$ are defined by
\begin{equation}
\chi(x,t)=\frac{\dot{\tau}}{2}\,x+\alpha(t), \quad D(x,t)=\frac{\ddot{\tau}}{8}\,x^2+\frac{\dot{\alpha}}{2}\,x+n(t)
\end{equation}
and the coefficients in the equation satisfy the determining equations
\begin{subequations}\label{det}
\begin{eqnarray}
\label{g1}&&\tau g_{1,t}+\chi\, g_{1,x}+(2A+\dot{\tau})g_1=0,\\
\label{g2}&&\tau g_{2,t}+\chi\, g_{2,x}+(2A+\dot{\tau})g_2=0,\\
\label{q1}&&\tau q_{1,t}+\chi\, q_{1,x}+(4A+\dot{\tau})q_1=0,\\
\label{q2}&&\tau q_{2,t}+\chi\, q_{2,x}+(4A+\dot{\tau})q_2=0,\\
\label{h1}&&\tau h_{1,t}+\chi\, h_{1,x}+\dot{\tau}h_1-D_t=0,\\
\label{h2}&&\tau h_{2,t}+\chi\, h_{2,x}+\dot{\tau}h_2+\dot{A}+\frac{\ddot{\tau}}{4}=0.
\end{eqnarray}
\end{subequations}
\end{prop}
\subsection{Symmetries for the equations with  constant coefficients}
Eqs. \eqref{det} are straightforward to solve in the special case where all the coefficients are  constants
$$ g(x,t)=g_1+ig_2,\quad q(x,t)=q_1+iq_2, \quad h(x,t)=h_1.$$
We sum up the results as

1.) $g\ne 0$, $q\ne 0$ (genuine cubic-quintic case):
The symmetry algebra is 4-dimensional and is spanned by
\begin{equation}\label{gal}
Q_1=\gen t,\quad Q_2=\gen x, \quad Q_3=t\gen x+\frac{x}{2}\gen \omega, \quad Q_4=\gen \omega.
\end{equation}
It is solvable and isomorphic to the 1-dimensional Galilei algebra $\gal(1)$.

2.) $q=0$, $g\ne 0$ (cubic case):
The symmetry algebra is 5-dimensional and has the basis
\begin{equation}\label{galsim}
\begin{split}
& Q_1=\gen t+h_1 \gen \omega,\quad Q_2=\gen x, \quad Q_3=t\gen x+\frac{x}{2}\gen \omega, \quad Q_4=\gen \omega,\\
& Q_5=\frac{x}{2}\gen x+t \gen t-\frac{1}{2}\rho \gen \rho+h_1 t \gen \omega.
\end{split}
\end{equation}
It is solvable and isomorphic to the 1-dimensional Galilei similitude algebra $\simi(1)\simeq \nil(1) \rhd \curl{Q_1, Q_5}$ with  $\nil(1)=\curl{Q_2, Q_3, Q_4}$ being the nilpotent ideal (Heisenberg algebra).

3.) $q\ne 0$, $g= 0$ (quintic case):
The symmetry algebra is 6-dimensional and is spanned by
\begin{equation}\label{sch}
\begin{split}
&Q_1=\gen t+h_1 \gen \omega,\quad Q_2=\gen x, \quad Q_3=t\gen x+\frac{x}{2}\gen \omega, \quad Q_4=\gen \omega,\\
&Q_5=\frac{x}{2}\gen x+t \gen t-\frac{1}{4}\rho \gen \rho+h_1 t \gen \omega,\quad Q_6=xt\gen x+t^2\gen t-\frac{1}{2}t\rho \gen \rho+(\frac{x^2}{4}+h_1 t^2)\gen \omega.
\end{split}
\end{equation}
It is non-solvable and isomorphic to the 1-dimensional Schr\"{o}dinger algebra $\sch(1)$ having the Levi decomposition
$$\sch(1)\simeq \curl{Q_1, Q_5, Q_6}\rhd \nil(1).$$ The simple algebra  $\curl{Q_1, Q_5, Q_6}$ is isomorphic to $\Sl(2,\mathbb{R})$.

The symmetry algebra allowed in the case  $h=h_1+ih_2$ with $h_2\neq 0$ is  four-dimensional and isomorphic to \eqref{gal} regardless of $g$ and $q$.

\subsection{One-dimensional algebras}

\begin{prop}\label{onedim}
The vector field \eqref{vf} can be transformed by the transformations \eqref{equivc} to one of the following canonical forms, in other words, there are precisely three inequivalent realizations:
\begin{equation}\label{canovf}
Q=\dw,\quad Q=\dt, \quad Q=\dx+m(t)\rho\dr.
\end{equation}
\end{prop}

\noindent Proof.
There are three different cases:

\noindent (i) $\tau(t)=\alpha(t)=0.$

We have $Q=A(t)\rho\dr+n(t)\dw$. Since
at least one of $g_1,g_2,q_1,q_2$ is different from zero, one of the four equations \eqref{g1}-\eqref{q2}  imposes $A(t)=0$, which means that the other three are satisfied identically. \eqref{h1} is satisfied if
$n(t)$ is a constant and \eqref{h2} automatically holds. Therefore $Q$ simplifies to
\begin{equation}\label{S11}
A_1^1:\quad Q=\dw
\end{equation}
as a canonical form of a one-dimensional algebra (pure gauge transformations). In this case
there  are no restrictions on the coefficients $g(x,t),q(x,t)$ and
$h(x,t)$.

\noindent (ii) $\tau(t)\neq0.$
We can transform $Q$ into $\gen t$ by choosing
$$\dot T=\tau^{-1},\quad \dot{\xi}=-\frac{\epsilon \alpha}{\tau^{3/2}},\quad R_0(t)=r_0
\exp\Big(\int\frac{A}{\tau}dt\Big), \quad \dot{\eta}=\frac{n}{\tau}-\frac{\alpha^2}{2\tau^2},$$ where $r_0$ is a constant, in the allowed transformation \eqref{equivc}. The canonical form of $Q$ (removing tildes) is
\begin{equation}\label{S12}
A_1^2: \quad Q=\dt.
\end{equation}
The corresponding equation contains coefficients depending only on $x$. $Q$ remains invariant under the transformation
\begin{equation}\label{e2}
\mathcal{E} :
\quad X=\epsilon x+\xi_0,
\quad T=t+T_0,
\quad \sigma=\frac{\rho}{r_0}
\quad \phi=\omega-\eta_0.
\end{equation}

\noindent (iii) $\tau(t)=0, \, \alpha(t)\neq0$.
 Similar arguments can be used to find the canonical vector field. The coefficients figuring in the invariant equation are
\begin{subequations}
\begin{eqnarray}
&g(x,t)=\big(g_1(t)+ig_2(t)\big)e^{-2xm(t)},\\
&q(x,t)=\big(q_1(t)+iq_2(t)\big)e^{-4xm(t)},\\
&h(x,t)=h_1(t)+i\big(h_2(t)-x\dot{m}(t)\big).
\end{eqnarray}
\end{subequations}
We can again use equivalence transformations to set $h_1(t)=0$, $h_2(t)=0$ and obtain
\begin{equation}\label{S13}
\begin{split}
A_1^3: \quad &Q=\dx+m(t)\rho\dr, \\
&g(x,t)=\big(g_1(t)+ig_2(t)\big)e^{-2xm(t)}\\
&q(x,t)=\big(q_1(t)+iq_2(t)\big)e^{-4xm(t)},\quad h(x,t)=-i\, x\, \dot{m}(t).
\end{split}
\end{equation}
$Q$ is invariant under the transformation
\begin{equation}\label{e3}
\mathcal{E} : \quad X=x+\xi_0, \quad T=t, \quad
\sigma=\frac{\rho}{r_0}, \quad \phi=\omega-\eta_0.
\end{equation}


\subsection{Low-dimensional algebras}
There are different classification methods in the literature. The strategy we shall pursuit here will be to adopt the dimensional approach which is basically based on using the classification results of the low-dimensional real Lie algebras up to dimension 4. We note that this approach is not effective for algebras beyond dimension 5 and preference  should  be given to the structural approach  to make the classification problem more tractable. In this approach one resorts to the known existing results on the structure of abstract Lie algebras.

The dimensional approach is an inductive one. Namely, we start from the already canonized one-dimensional algebras and enlarge them to two-dimensional ones by adding a second generator and require them to form a Lie algebra. We then use the equivalence transformations to simplify the obtained algebra. Invariance under this algebra will of course restrict the coefficients in the equation. We shall carry out this procedure up to dimension 4 until the corresponding invariant equations contain either functions of a single variable or constants only. Beyond $\dim L=4$  it looks reasonable to check whether the determining equations can possibly allow any further extensions that would automatically reveal  maximal symmetry algebras. This is done by inserting the coefficients of the 4-dimensional symmetry algebras and the symmetry vector fields into \eqref{det} and then examining the compatibility of the parameters or solving for the functions in the equations invariant under 4-dimensional algebras.

Throughout  we shall use the nomenclature of the Lie algebra classification given for example in \cite{Basarab-Horwath2001} and list only maximal symmetry algebras.

From  Proposition \ref{onedim} we know that Eq. \eqref{canonic} admits the maximal symmetry group (the pure gauge transformations) generated by $Q_1=\gen \omega$ for any possible choice of the coefficients. The symmetry algebras larger than dimension one will naturally be extensions of $Q_1=\gen \omega$ to higher dimensions. So  every Lie symmetry algebra $L$ with dimension $\dim L\geq 2$ contains $Q_1=\gen \omega$ as a subalgebra (actually the center).

\subsubsection{Abelian algebras}
If we let $Q_1=\dw$, $Q_2=Q$
of the form \eqref{vf} and impose the condition
$[Q_1,Q]=0$. It is satisfied without any restriction
on the form of $Q$.  Simplifying by equivalence transformations we obtain
\begin{equation}\label{S21}
\begin{split}
A^1_{2.1}: \quad &Q_1=\dw, \quad  Q_2=\dt, \\
&g(x,t)=g_1(x)+ig_2(x),\\
&q(x,t)=q_1(x)+iq_2(x), \quad h(x,t)=h_1(x)+ih_2(x).
\end{split}
\end{equation}
\begin{equation}\label{S22}
\begin{split}
A^2_{2.1}:
\quad
&Q_1=\partial_{\omega},
\quad  Q_2=\dx+m(t)\rho\dr, \\
&g(x,t)=\big(g_1(t)+ig_2(t)\big)e^{-2xm(t)},\\
&q(x,t)=\big(q_1(t)+iq_2(t)\big)e^{-4xm(t)},\quad
h(x,t)=-i\, x\, \dot{m}(t).
\end{split}
\end{equation}

$Q_1=\dw$ commutes with the general element $Q$ so that we cannot obtain a two dimensional non-abelian  algebra.

\subsection{Three-dimensional Algebras}
A real 3-dimensional algebra is either simple or solvable.

\subsubsection{Simple algebras}
The only 3-dimensional simple algebras are $\Sl(2,\mathbb{R})$ and $\So(3,\mathbb{R})$.
The first algebra contains a 2-dimensional non-abelian algebra. This implies that there can be no $\Sl(2,\mathbb{R})$ realizations. We also found that  $\So(3,\mathbb{R})$ can not be realized in terms of vector fields \eqref{vf}.

\subsubsection{Solvable algebras}
All 3-dimensional solvable algebras have 2-dimensional abelian ideals (nilradicals). We choose a basis $\curl{Q_1, Q_2, Q_3}$ having $\curl{Q_1, Q_2}$ in the ideal and impose the commutation relations
\begin{equation}\label{3dcomm}
[Q_1,Q_3]=0, \quad
[Q_2,Q_3]=a_1Q_1+a_2Q_2.
\end{equation}
If $a_2\neq 0$, by a change of basis one can always set $a_1=0$, $a_2=1$. This is the case of decomposable algebra $A_{3.2}$. If $a_2=0$, then we can have $a_1=0$, which is the case of abelian algebra $A_{3.1}$ or we have $a_1=1$, and that corresponds to the nilpotent algebra $A_{3.5}$.

Once  the form of $Q_3$ has been found from the commutator relations \eqref{3dcomm}, the allowed transformations leaving the ideal $\curl{Q_1, Q_2}$ invariant (the residual equivalence group) are then used to simplify $Q_3$.

\subsubsection{Abelian case}
We start with  $A^1_{2.1}$. Let $Q_1=\dw$, $Q_2=\dt$, $Q_3=Q$.  We already have $[Q_1,Q]=0$. We then impose the condition
\begin{equation}
[Q_2,Q]=(\dot{\alpha}+\frac{x}{2}\ddot{\tau})\dx+\dot{\tau}\dt+\dot{A}\rho\dr+(\dot{n}+\frac{x}{2}\ddot{\alpha}+\frac{x^2}{8}\dddot{\tau})\dw=0.
\end{equation}
We must have $\tau(t)=\tau_0$, $\alpha(t)=\alpha_0$, $n(t)=n_0$, all of which are constants. This gives the generator $Q=\alpha_0\dx+\tau_0\dt+A_0\rho\dr+n_0\dw$. By a change of basis we can make $\tau_0\rightarrow 0.$ If $\alpha_0=0$, one of the equations \eqref{g1}-\eqref{q2} requires
$A_0=0$. So assume that $\alpha_0\neq 0$. We can rescale $Q$ to have $\alpha_0\rightarrow 1$ and so that $Q=\dx+A_0\rho\dr+n_0\dw$. We solve the determining equations for the coefficients and find after relabeling the constants
\begin{equation}
\begin{split}\label{a214}
&Q_1=\dw,
\quad
Q_2=\dt, \quad Q_3=\dx+a\rho\dr, \\
&g(x,t)=(g_1+ig_2)\exp(-2ax),\\
&q(x,t)=(q_1+iq_2)\exp(-4ax), \\
&h(x,t)=h_1+ih_2, \qquad
a,g_1,g_2,q_1,q_2,h_1,h_2 \in\mathbb{R}.
\end{split}
\end{equation}
It is not possible to kill $h_1$ or $h_2$ through the allowed transformations \eqref{e2}.

Now let us take  $A^2_{2.1}$ as the ideal. Set $Q_1=\dw$, $Q_2=\dx+m(t)\rho\dr$, $Q_3=Q$. We have
\begin{equation}
[Q_2,Q]=\frac{\dot{\alpha}}{2}\dx-\tau\dot{m}\rho\dr+(\frac{\dot{\alpha}}{2}+\frac{x}{4}\ddot{\tau})\dw=0.
\end{equation}
(i) If $\tau(t)=0$, then $\alpha(t)=\alpha_0$ and we have $Q=\alpha_0\dx+A(t)\rho\dr+n(t)\dw$. We change the basis to put $\alpha_0=0$ ($A(t)$ is not the same as before, but otherwise arbitrary). Solving the determining equations we have  ${Q}=\dw$ and there is no extension. (ii) Assume that $\tau(t)\neq 0$. Then we must have $m(t)=m_0$, $\alpha(t)=\alpha_0$, $\tau(t)=\tau_0$.
This means we have $Q_2=\dx+m_0\rho\dr$, $Q=\alpha_0\dx+\tau_0\dt+A(t)\rho\dr+n(t)\dw$. We can rescale $Q$ to have $\tau_0\rightarrow 1$. A change of basis gives ${Q}=\dt+A(t)\rho \dr+n(t)\dw$. When we solve the determining equations, we see that $A(t)$ and $n(t)$ must be constants.
We find the abelian  algebra
\begin{equation}\label{S24}
\begin{split}
A^1_{3.1}:
\quad
&Q_1=\dw, \quad Q_2=\dx+a\rho\dr,
\quad
Q_3=\dt+b\rho\dr, \\
&g(x,t)=(g_1+ig_2)\exp[-2(ax+bt)],\\
&q(x,t)=(q_1+iq_2)\exp[-4(ax+bt)], \\
&h(x,t)=0, \qquad
a,b,g_1,g_2,q_1,q_2 \in\mathbb{R}.
\end{split}
\end{equation}
Note that \eqref{a214} is not included in the canonical list of three-dimensional abelian algebras as it  is equivalent to the algebra $A^1_{3.1}$ through the transformation
\begin{equation}
X=x+\xi_0, \quad T=t+T_0, \quad \sigma=\rho \exp(h_2t), \quad \phi=\omega-h_1t+\eta_0.
\end{equation}
\subsubsection{Decomposable case}
We start with $A^1_{2.1}$ and   let $Q_1=\dw$, $Q_2=\dt$, $Q_3=Q$ and we impose
\begin{equation}
[Q_2,Q]=(\dot{\alpha}+\frac{x}{2}\ddot{\tau})\dx+\dot{\tau}\dt+\dot{A}\rho\dr+(\dot{n}+\frac{x}{2}\ddot{\alpha}+\frac{x^2}{8}\dddot{\tau})\dw=Q_2.
\end{equation}
We find
\begin{equation}
Q=(\alpha_0+\frac{x}{2})\dx+(t+\tau_0)\dt+A_0\rho\dr+n_0\dw
\end{equation}
where all the parameters are constant. We can assume $\tau_0=0$ up to a change of basis $Q\rightarrow Q-\tau_0 Q_2$. Applying the allowed transformation
\eqref{e2} with the choice $\epsilon=1$, $\xi_0=2\alpha_0$, $T_0=0$ and solving the  determining equations we obtain
\begin{equation}\label{a122}
\begin{split}
A^1_{3.2}:
\quad
&Q_1=\dw, \quad Q_2=\dt, \quad Q_3=\frac{x}{2}\,\dx+t\dt+a\rho\dr,\\
&g(x,t)=\frac{g_1+ig_2}{x^{2(2a+1)}}, \quad
q(x,t)=\frac{q_1+iq_2}{x^{2(4a+1)}}, \quad
h(x,t)=\frac{h_1+ih_2}{x^2},
\end{split}
\end{equation}
with some constants $a,g_1,g_2,h_1,h_2\in\mathbb{R}$.

If we continue with  $A^2_{2.1}$, for $Q_1=\dw$, $Q_2=\dx+m(t)\rho\dr$ and $Q_3=Q$,
the commutation condition
\begin{equation}
[Q_2,Q]=\frac{\dot{\tau}}{2}\dx-\tau \dot{m}\rho\dr+(\frac{\dot{\alpha}}{2}+\frac{x}{4}\ddot{\tau})\dw=Q_2
\end{equation}
requires  $\tau(t)=2(t+\tau_0)$, $m(t)=\frac{m_0}{\sqrt{t+\tau_0}}$, $\alpha(t)=\alpha_0$, which means  we have
$$Q_2=\dx+\frac{m_0}{\sqrt{t+\tau_0}}\rho\dr,\quad Q=(x+\alpha_0)\dx+2(t+\tau_0)\dt+A(t)\rho\dr+n(t)\dw.$$ The allowed transformation \eqref{e3}
transforms away $\alpha_0$. We solve the determining equations and find that $A(t)$ and $n(t)$ must be constants.
\begin{equation}\label{S26}
\begin{split}
A^2_{3.2}: \quad
&Q_1=\dw, \quad Q_2=\dx+\frac{a}{\sqrt{t+b}}\,\rho\,\dr,\quad
Q_3=x\dx+2(t+b)\dt+c\rho\,\dr,  \\
&g(x,t)=\frac{g_1+ig_2}{(t+b)^{1+c}}\exp\Big(\frac{-2ax}{\sqrt{t+b}}\Big)\\
&q(x,t)=\frac{q_1+iq_2}{(t+b)^{1+2c}}\exp\Big(\frac{-4ax}{\sqrt{t+b}}\Big), \quad h(x,t)=i\frac{ax}{2(t+b)^{3/2}}
\end{split}
\end{equation}
with some constants $a,b,c,g_1,g_2,q_1,q_2$.
\subsubsection{Nilpotent algebras}

We try to realize $A_{3.5}$ by extending the two dimensional
abelian algebras $A_{2.1}=\{Q_1,Q_2\}$ with an element $Q_3=Q$
that will satisfy $[Q_2,Q_3]=Q_1$. We give the final result skipping the details. $A^1_{2.1}$ leads to the
nilpotent algebra
\begin{equation}
\begin{split}
A^1_{3.5}: \quad &Q_1=\dw, \quad  Q_2=\dt, \quad Q_3=\dx+a\rho\dr+t\dw, \\
&g(x,t)=(g_1+ig_2)\exp(-2ax), \quad q(x,t)=(q_1+iq_2)\exp(-4ax), \\
&h(x,t)=x+h_1+ih_2, \qquad g_1, g_2,q_1,q_2
h_1, h_2 \in \mathbb{R}.
\end{split}
\end{equation}
We find two different realizations from $A^2_{2.1}$.
\begin{equation}
\begin{split}
A^2_{3.5}: \quad &Q_1=\partial_{\omega}, \quad
Q_2=\dx+a\rho\dr, \quad
                 Q_3=2t\dx+\tau_0\dt+b\rho\dr+x\dw,\\
                 &g(x,t)=(g_1+ig_2)\exp[-2ax+\frac{2}{\tau_0}(at^2-bt)],\\
                 &q(x,t)=(q_1+iq_2)\exp[-4ax+\frac{4}{\tau_0}(at^2-bt)],\quad h(x,t)=0
\end{split}
\end{equation}
with some constants $\tau_0\neq0, a, b, g_1,g_2,q_1,q_2 \in \mathbb{R}$.
If $\tau_0=0$ we have
\begin{equation}
\begin{split}
A^3_{3.5}: \quad &Q_1=\partial_{\omega}, \quad
Q_2=\dx, \quad
                 Q_3=2t\dx+x\dw,\\
                 &g(x,t)=g_1(t)+ig_2(t), \quad q(x,t)=q_1(t)+iq_2(t), \quad
                 h(x,t)=0.
\end{split}
\end{equation}
This completes the analysis of three dimensional algebras.
\subsection{Four-dimensional algebras}

For the canonical list of four-dimensional Lie algebras we refer
the reader to \cite{Basarab-Horwath2001}. In the following we skip the details and list only the algebras and the equations.

\paragraph{Non-solvable algebras}
$A_{3.3}$ trivially extends to the algebra (isomorphic to $\gl(2,\mathbb{R})$)
\begin{equation}\label{S41}
\begin{split}
A_{3.3}\oplus A_1: \quad &Q_1=\dt, \quad Q_2=\frac{x}{2}\,\dx+t\dt-\frac{1}{4}\,\rho\,\dr, \\
               &Q_3=xt\dx+t^2\dt-\frac{1}{2}\,t\,\rho\,\dr+\frac{x^2}{4}\dw, \quad Q_4=\dw,\\
&g(x,t)=(g_1+ig_2)\,x^{-1}, \quad q(x,t)=q_1+iq_2, \quad h(x,t)=(h_1+ih_2)x^{-2}.
\end{split}
\end{equation}
\paragraph{Nilpotent algebras}
\begin{equation}\label{S43}
\begin{split}
A_{4.1}: \quad &Q_1=\dw, \quad
                Q_2=\dx, \quad Q_3=\dt+b\rho\dr, \quad Q_4=t\dx+\frac{x}{2}\dw,\\
&g(x,t)=(g_1+ig_2)\exp(-2bt), \quad q(x,t)=(q_1+iq_2)\exp(-4bt), \\
&h(x,t)=0, \quad
b,g_1,g_2,q_1,q_2 \in\mathbb{R}
\end{split}
\end{equation}
 is the semi-direct sum of the three-dimensional abelian
ideal $A_{3.1}$ with the one-dimensional algebra that generates
Galilean boosts.

\paragraph{Indecomposable solvable algebras}
\begin{equation}\label{S44}
\begin{split}
A_{4.8}: \quad &Q_1=\partial_{\omega}, \quad
Q_2=\dx, \quad Q_3=2t\dx+x\dw, \quad Q_4=x\dx+2t\dt+a\rho\dr,\\
                 &g(x,t)=\frac{g_1+ig_2}{t^{1+a}}, \quad
                  q(x,t)=\frac{q_1+iq_2}{t^{1+2a}}, \quad h(x,t)=0.
\end{split}
\end{equation}

\begin{equation}\label{S45}
\begin{split}
A_{4.9}: \quad &Q_1=\partial_{\omega}, \quad
Q_2=\dx, \quad   Q_3=2t\dx+x\dw, \quad \\
                 &Q_4=xt\dx+(1+t^2)\dt-\frac{1}{2}(b+t)\rho\dr+\frac{x^2}{4}\dw,\\
                 &g(x,t)=\frac{g_1+ig_2}{\sqrt{1+t^2}}\exp\big(b\arctan t\big), \\
                 &q(x,t)=(q_1+iq_2)\exp\big(2b\arctan t\big), \quad
                 h(x,t)=0.
\end{split}
\end{equation}

\subsection{Five and six-dimensional algebras}

We directly solve the determining equations  for the  classes \eqref{S41}-\eqref{S45}  to
find their full symmetry groups.
\subsubsection{The algebra $A_{3.3}\oplus A_1$}
From eqs. \eqref{g1}-\eqref{h2} we find
\begin{subequations}
\begin{eqnarray}
\Big(\frac{4A+\dot{\tau}}{2x}-\frac{\alpha}{x^2}\Big)g_1&=&0, \label{511}\\
\Big(\frac{4A+\dot{\tau}}{2x}-\frac{\alpha}{x^2}\Big)g_2&=&0,\label{512}\\
(4A+\dot{\tau})q_1&=&0,\label{513}\\
(4A+\dot{\tau})q_2&=&0,\label{514}\\
\frac{2\alpha}{x^3}h_1+\dot{n}+\frac{\ddot{\alpha}}{2}x+\frac{\dddot{\tau}}{8}x^2&=&0\label{515}\\
\frac{2\alpha}{x^3}h_2-\dot{A}-\frac{\ddot{\tau}}{4}&=&0.   \label{516}
\end{eqnarray}
\end{subequations}
We distinguish between two cases. (i) Let $\alpha(t)\neq 0$. Then we must have $g_1=g_2=h_1=h_2=0$, which means this is the constant coefficient quintic case with zero potential.
From  \eqref{515} we see that
\begin{equation}
\tau(t)=c_1+c_2t+c_3t^2, \quad \alpha(t)=c_4t+c_5, \quad n(t)=c_6
\end{equation}
with  constants $c_i$, $i=1,..,6.$ Since $q_1$ and $q_2$ cannot be both zero, \eqref{513} or \eqref{514} requires $A(t)=-{\dot{\tau}}/{4}=-({c_2}+2{c_3}t)/4$ and \eqref{516} is satisfied. The vector field $Q$ in \eqref{vf} has six arbitrary constants. Therefore, we end up with a six-dimensional Lie algebra
\begin{equation}\label{6}
\begin{split}
\quad \quad &Q_1=\dt, \quad Q_2=\frac{x}{2}\,\dx+t\dt-\frac{1}{4}\,\rho\,\dr,\quad  Q_3=xt\dx+t^2\dt-\frac{1}{2}\,t\,\rho\,\dr+\frac{x^2}{4}\dw,\\
            & Q_4=\dw, \quad Q_5=t\dx+\frac{x}{2}\dw, \quad Q_6=\dx,\\
&g(x,t)=0, \quad q(x,t)=q_1+iq_2, \quad h(x,t)=0, \quad
q_1,q_2\in\mathbb{R}.
\end{split}
\end{equation}
(ii) If $\alpha(t)=0$, under the condition that one of $g_1,g_2,q_1,q_2$ is not zero we must have $4A+\dot{\tau}=0$ and the remaining part of the analysis is similar to (i) but in this case we have $c_4=c_5=0$. There are  four arbitrary constants $c_1$, $c_2$, $c_3$, $c_6$ in $Q$.  Therefore, it does not allow any further extension.
\subsubsection{The algebra $A_{4.1}$}
In this case we have the following set of determining equations
\begin{subequations}
\begin{eqnarray}
(-2b\tau+2A+\dot{\tau}) g_1&=&0, \label{541}\\
(-2b\tau+2A+\dot{\tau}) g_2&=&0, \label{542}\\
(-4b\tau+4A+\dot{\tau}) q_1&=&0, \label{543}\\
(-4b\tau+4A+\dot{\tau}) q_2&=&0, \label{544}\\
\dot{n}+\frac{\ddot{\alpha}}{2}x+\frac{\dddot{\tau}}{8}x^2&=&0\label{545},\\
\dot{A}+\frac{\ddot{\tau}}{4}&=&0.   \label{546}
\end{eqnarray}
\end{subequations}
From \eqref{545} it immediately follows that $\tau(t)=c_1+c_2t+c_3t^2$, $\alpha(t)=c_4 t+c_5$, $n(t)=c_6$. From now on we are going to distinguish between three cases.

(i) Let $q_1=q_2=0$, $g_1^2+g_2^2\neq 0$. We have from one of \eqref{541} and \eqref{542} that $A=b\tau-\frac{\dot{\tau}}{2}$. Using this in \eqref{546} results in the condition
\begin{equation}\label{bt}
bc_2-\frac{c_3}{2}+2bc_3t=0.
\end{equation}
If $b=0$, we only have the condition $c_3=0$, hence we have five arbitrary constants in the vector field leading to a five-dimensional Lie algebra
\begin{equation}\label{5}
\begin{split}
&Q_1=\dx, \quad Q_2=\dt, \quad
Q_3=\frac{x}{2}\,\dx+t\dt-\frac{1}{2}\rho\dr, \\
&Q_4=\dw, \quad Q_5=t\dx+\frac{x}{2}\dw,\\
&g(x,t)=g_1+ig_2, \quad q(x,t)=0, \quad h(x,t)=0, \quad
g_1,g_2\in\mathbb{R}.
\end{split}
\end{equation}Let  $b\neq 0$.
\eqref{bt} then forces $c_2=c_3=0$, therefore the algebra remains as a four-dimensional one.

(ii) The case $g_1=g_2=0$, $q_1^2+q_2^2\neq 0$.  We have $A=b\tau-\frac{\dot{\tau}}{4}$ from one of \eqref{543} and \eqref{544}, whereas \eqref{546} gives
\begin{equation}\label{bt2}
b(c_2+2c_3t)=0.
\end{equation}
Similar to the previous case, the case  $b=0$ corresponds to the six-dimensional algebra given in \eqref{6}. If $b\neq 0$, we again need to have $c_2=c_3=0$ from \eqref{bt2}, which means that the four-dimensional algebra  cannot be extended.

(iii) $g_1^2+g_2^2\neq 0$,  $q_1^2+q_2^2\neq 0$. From \eqref{541}-\eqref{544} we obtain $c_2=c_3=0$ and $A(t)=bc_1$. Thus the dimension does not exceed four in the cubic-quintic case, either.

\subsubsection{The algebra $A_{4.8}$}
We find the set of determining equations as
\begin{subequations}
\begin{eqnarray}
\Big(-\frac{1+a}{t}\,\tau+2A+\dot{\tau}\Big) g_1&=&0, \label{551}\\
\Big(-\frac{1+a}{t}\,\tau+2A+\dot{\tau}\Big) g_2&=&0, \label{552}\\
\Big(-\frac{1+2a}{t}\,\tau+4A+\dot{\tau}\Big) q_1&=&0, \label{553}\\
\Big(-\frac{1+2a}{t}\,\tau+4A+\dot{\tau}\Big) q_2&=&0, \label{554}\\
\dot{n}+\frac{\ddot{\alpha}}{2}x+\frac{\dddot{\tau}}{8}x^2&=&0\label{555},\\
\dot{A}+\frac{\ddot{\tau}}{4}&=&0.   \label{556}
\end{eqnarray}
\end{subequations}
We have $\tau(t)=c_1+c_2t+c_3t^2$, $\alpha(t)=c_4 t+c_5$, $n(t)=c_6$ from \eqref{555}.

 (i) $q_1=q_2=0$, $g_1^2+g_2^2\neq 0$.
(A) The case $a=-1$. We need to have $c_3=0$. There are five arbitrary constants in the vector field. This is exactly the algebra in \eqref{5}.

\noindent (B) The case $a=0$.  We have $c_1=0$. This gives again  a five-dimensional Lie algebra
\begin{equation}
\begin{split}\label{5C}
&Q_1=\dx, \quad  Q_2=xt\dx+t^2\dt-\frac{1}{2}\,t\,\rho\,\dr+\frac{x^2}{4}\dw, \\
& Q_3=x\dx+2t\dt, \quad Q_4=\partial_{\omega}, \quad Q_5=2t\dx+x\dw, \\
&g(x,t)=(g_1+ig_2)\,t^{-1}, \quad
                  q(x,t)=0, \quad h(x,t)=0,
\end{split}
\end{equation}
which is actually  equivalent to \eqref{5}.

(ii) $g_1=g_2=0$, $q_1^2+q_2^2\neq 0$. When we solve $A(t)$ from one of \eqref{553} or \eqref{554} and plug it in \eqref{556}, we see that
\begin{equation}
(1+2a)c_1=0, \quad (1+2a)c_3=0
\end{equation}
must hold. The case $a=-\frac{1}{2}$ corresponds to the six-dimensional algebra \eqref{6}. If $a\neq -\frac{1}{2}$, then we have $c_1=c_3=0$, which means that the dimension of the algebra cannot exceed four.

(iii) $g_1^2+g_2^2\neq 0$,  $q_1^2+q_2^2\neq 0$. From \eqref{551}-\eqref{554} we get $c_1=c_3=0$ and $A(t)={(ac_2)}/{2}$. \eqref{556} holds in this case and we have a four-dimensional algebra.
\subsubsection{The algebra $A_{4.9}$}
The set of determining equations are
\begin{subequations}
\begin{eqnarray}
\Big(\frac{b-t}{1+t^2}\,\tau+2A+\dot{\tau}\Big) g_1&=&0, \label{561}\\
\Big(\frac{b-t}{1+t^2}\,\tau+2A+\dot{\tau}\Big) g_2&=&0, \label{562}\\
\Big(\frac{2b}{1+t^2}\,\tau+4A+\dot{\tau}\Big) q_1&=&0, \label{563}\\
\Big(\frac{2b}{1+t^2}\,\tau+4A+\dot{\tau}\Big) q_2&=&0, \label{564}\\
\dot{n}+\frac{\ddot{\alpha}}{2}x+\frac{\dddot{\tau}}{8}x^2&=&0\label{565},\\
\dot{A}+\frac{\ddot{\tau}}{4}&=&0.   \label{566}
\end{eqnarray}
\end{subequations}
We have $\tau(t)=c_1+c_2t+c_3t^2$, $\alpha(t)=c_4 t+c_5$, $n(t)=c_6$ from \eqref{565}.

(i) $q_1=q_2=0$, $g_1^2+g_2^2\neq 0$. Solving $A(t)$ from one of \eqref{561} or \eqref{562} results in
$A(t)=\frac{t-b}{2(1+t^2)}\,\tau-\frac{\dot{\tau}}{2}$. From \eqref{566} we have
\begin{equation}
-(c_3+bc_2-c_1)+2(bc_1+c_2-bc_3)t+(c_3+bc_2-c_1)t^2=0.
\end{equation}
Coefficients of $t^0$ and $t^2$ require $c_1=c_3+bc_2$, and coefficient of $t$ gives $c_2=0$ independently of $b$. We have two constraints on six arbitrary constants, therefore, the dimension of the algebra is four and  in this case $A_{4.9}$ cannot allow a larger algebra.

(ii) $g_1=g_2=0$, $q_1^2+q_2^2\neq 0$. We obtain  $A(t)=-\frac{b}{2(1+t^2)}\,\tau-\frac{\dot{\tau}}{4}$ from one of \eqref{563} and \eqref{564} and
\begin{equation}
b\Big(c_2+2(c_3-c_1)t-c_2t^2\Big)=0
\end{equation}
from \eqref{566}.
We already have the result for $b=0$, the constant-coefficient case in \eqref{6}. For $b\neq 0$ we need to have $c_2=0$, $c_1=c_3$ and thus the algebra cannot be enlarged to a higher dimension.

(iii) The cubic-quintic case also does not allow a larger algebra.

We now sum up our results.

\begin{thm}
The symmetry group of the genuine (g and q not both zero) variable coefficient CQNLS equation can be at most 4-dimensional. There are precisely four inequivalent classes of equations given by \eqref{S41}, \eqref{S43}, \eqref{S44}, \eqref{S45}.
\end{thm}

\begin{thm}
Any variable coefficient CQNLS equation with a 5- or 6-dimensional symmetry group can be transformed into the standard cubic equation \eqref{5} or quintic equation \eqref{6}.

The symmetry algebra is isomorphic to the 1-dimensional Galilei similitude  algebra $\simi(1)$ in the first case, and to the 1-dimensional Schr\"{o}dinger $\sch(1)$
algebra in the second case.
\end{thm}

\section{Analogy with the standard CQNLS equation}
We can take advantage of the equivalence transformations to establish the conditions  for the transformability of equation \eqref{cqsch} with $f=1$ to the constant-coefficient cubic-quintic equation, namely
\begin{equation}\label{tkosul}
 \tilde{f}=1, \quad \tilde{g}=a_1+ia_2\neq 0, \quad \tilde{q}=b_1+ib_2\neq 0,
\quad \tilde{k}=\tilde{h}=0,
\end{equation}
where $a_1,a_2,b_1,b_2$ are real  constants.

We already know that
$\tilde{f}=1$ constrains $X$ to
\begin{equation}\label{X}
X(x,t)=\epsilon\sqrt{\dot{T}}x+\xi(t).
\end{equation}
\eqref{g} and \eqref{q} imply that
\begin{subequations}\label{gtqt}
\begin{eqnarray}
\tilde{g}=(g_1+ig_2)\frac{R^2}{\dot{T}}=a_1+ia_2,\\
\tilde{q}=(q_1+iq_2)\frac{R^4}{\dot{T}}=b_1+ib_2.
\end{eqnarray}
\end{subequations}
Let us assume that $g_1(x,t)\neq 0$, and this requires $a_1\neq 0$. From \eqref{gtqt} we have  the requirements
\begin{equation}\label{gq}
g(x,t)=g_1(x,t)\,(1+i\frac{a_2}{a_1}), \quad q(x,t)=\frac{b_1+ib_2}{a_1^2}\frac{g_1^2(x,t)}{\gamma(t)}.
\end{equation}
We further have
\begin{equation}\label{TR}
T(t)=\int\gamma(t) dt, \quad R(x,t)=\Big(\frac{a_1\gamma(t)}{g_1(x,t)}\Big)^{1/2}.
\end{equation}
The condition $\tilde{k}=0$ gives
\begin{equation}
X_x(2\frac{R_x}{R}+k_1)+i(X_t+2X_x\theta_x+k_2 X_x)=0
\end{equation}
leading to
\begin{equation}\label{k1}
k_1(x,t)=\frac{g_{1,x}}{g_1},
\end{equation}
\begin{equation}\label{teta}
\theta(x,t)=-\frac{\dot{\gamma}}{8\gamma}\,x^2-\frac{\dot{\xi}}{2\epsilon\sqrt{\gamma}}\, x-\frac{1}{2}\int k_2(x,t) \,dx+\eta(t).
\end{equation}
Having determined the transformations, we can construct the corresponding admissible potentials  from the condition $\tilde{h}=0$ as
\begin{subequations}\label{h1h2}
\begin{eqnarray}
\nonumber h_1(x,t)&=&\Big(\frac{3}{16}\big(\frac{\dot{\gamma}}{\gamma}\big)^2-\frac{\ddot{\gamma}}{8\gamma}\Big)\,x^2+\frac{1}{2\epsilon}\Big(\frac{\dot{\xi}\dot{\gamma}-\ddot{\xi}\gamma}{ \gamma^{3/2}}\Big)\, x\\
&+&\frac{g_{1,xx}}{2g_1}-\frac{g_{1,x}^2}{4g_1^2}-\frac{k_2^2}{4}-\frac{1}{2}\int k_{2,t}\, dx+\frac{\dot{\xi}^2}{4\gamma}+\dot{\eta},\label{H1}\\
h_2(x,t)&=&\frac{g_{1,t}}{2g_1}+\frac{k_2g_{1,x}}{2g_1}+\frac{1}{2}k_{2,x}-\frac{\dot{\gamma}}{4\gamma}.
\end{eqnarray}
\end{subequations}
Summarizing,  transformation of \eqref{cqsch} with $f=1$ and the coefficients given by \eqref{gq}, \eqref{k1} and \eqref{h1h2} to the standard cubic-quintic equation can be made possible by the special equivalence group given by $X,T,R,\theta$ in \eqref{X}, \eqref{TR}, \eqref{teta}.

\begin{rmk} Radial case:

For $k(x,t)=\frac{n-1}{x}$, from \eqref{k1} we must have $g_1(x,t)=G(t)x^{n-1}$ with an arbitrary $G$. Provided the condition  \eqref{gq}  holds, the admissible potential
\begin{subequations}\label{h1h2a}
\begin{eqnarray}
\nonumber h_1(x,t)&=&\Big(\frac{3}{16}\big(\frac{\dot{\gamma}}{\gamma}\big)^2-\frac{\ddot{\gamma}}{8\gamma}\Big)\,x^2+
\frac{1}{2\epsilon}\Big(\frac{\dot{\xi}\dot{\gamma}-\ddot{\xi}\gamma}{ \gamma^{3/2}}\Big)\, x\\
&+&\frac{(n-1)(n-3)}{4x^2}+\frac{1}{4}\frac{\dot{\xi}^2}{\gamma}+\dot{\eta},\label{HH1}\\
h_2(x,t)&=&\frac{1}{4}\big(2\frac{\dot{G}}{G}-\frac{\dot{\gamma}}{\gamma}\big),
\end{eqnarray}
\end{subequations}
allows the transformation given by  \eqref{X}, \eqref{TR} and \eqref{teta}.
\end{rmk}
The coefficient of the quadratic term in $h_1$ can be identified  with a multiple of  the Schwarzian derivative of $T(t)$ in the form  $-\{T;t\}/8$. This implies  that if the potential does not contain a quadratic term, then the time transformation $T$ appears to be the  linear fractional (or M\"{o}bius) transformations in $t$ depending on three  parameters. However, $\gamma$ in $q$ of \eqref{gq} is fixed by the relation $\gamma=T'(t)$ in this case.
\begin{rmk} In the cubic-quintic case, four-dimensional algebras  $A_{3.3}\oplus A_1$, $A_{4.8}$ and $A_{4.9}$  fail to satisfy all of the conditions \eqref{gq}, \eqref{k1}, \eqref{h1h2} and therefore cannot be transformed to the cubic-quintic equation with constant coefficients. These conditions hold for $A_{4.1}$ only when $b=0$, which  is of course trivial.
\end{rmk}


\end{document}